\documentclass{ws-p9-75x6-50s}

\begin{document}

\title{Finite temperature dynamics\\ near quantum phase transitions}

\author{Subir Sachdev}

\address{Department of Physics, Yale University,\\ P.O. Box 208120, New
Haven, CT 06520-8120, USA.\\E-mail: subir.sachdev@yale.edu}




\maketitle

\abstracts{We review the non-zero temperature relaxational
dynamics of quantum systems near a zero temperature, second-order
phase transition. We begin with the quantum Ising chain, for which
universal and exact results for the relaxation rates can be
obtained in all the distinct limiting regimes of the phase
diagram. Next, we describe the crossovers in the electron spectral
function near a transition involving a change in the pairing
symmetry of BCS superconductors in two dimensions. Finally, we
consider dynamic spin models which may provide a mean-field
description of magnetic ordering transitions in the heavy fermion
compounds. }

\begin{center}
{\tt Keynote talk at the 11th International Conference on\\ Recent
Progress in Many-Body Theories,\\ UMIST, Manchester UK, 9-13 July,
2001}
\end{center}

\section{Introduction}

The description of the long-time, low temperature, collective
dynamics of condensed matter systems is one of the central aims of
quantum many body theory\cite{forster,lovesey,landau}. Numerous
successful theories have been developed for a variety of
materials, and we can broadly separate them into two (not entirely
distinct) categories:
\newline
{\em Order parameter dynamics:} Often the system is in or near a
phase with a broken symmetry. This can be used to define an order
parameter which, along with globally conserved quantities, obey
equations of motion which define the collective, long-time
dynamics. The simplest example of this was the van Hove theory,
which evolved into a sophisticated description of the highly
non-trivial dynamics near finite temperature second order phase
transitions~\cite{halphoh}.
\newline
{\em Quasiparticle dynamics:} The elementary excitations above a
particular ground states are identified, and a transport equation
for their collision and interactions provides a description of the
low frequency response functions. Landau's Fermi liquid
theory~\cite{landau} is the familiar example of such a theory.
\newline
One common feature of both classes of models is that the dynamic
equations controlling the lowest frequency dynamics are ultimately
{\em classical}. They do, nevertheless, provide a description of
quantum systems at low temperature. Quantum mechanics does play a
fundamental role in determining the nature of the ground state,
its excitations, and of effective coupling constants in the
equations of motion, but the equations are finally expressed in
terms of classical, collective degrees of freedom. For the order
parameter dynamics, the justification for a classical description
is that the typical relaxation frequency is smaller than $k_B
T/\hbar$ (where $T$ is the absolute temperature), while in the
quasiparticle models we require a collision frequency smaller than
$k_B T/\hbar$.

In this paper, we will review recent studies of finite temperature
dynamics near second-order quantum phase transitions\cite{book}.
We will find that the two classes of classical models discussed
above do apply over a significant portion of the phase diagram.
However, we will also discuss the novel {\em quantum critical}
region\cite{chn}, where both descriptions are inadequate. The
fundamental property of this region\cite{sy,csy} is that the
characteristic frequency is a number of order unity (which is
usually universal) times $k_B T/\hbar$: so neither classical
description is evidently applicable. A theory of quantum-critical
dynamics can, of course, be obtained from an exact solution of the
critical field theory describing the phase transition: we will see
an example of this in Section~\ref{sec:ising} and the results
yield considerable insight. However, such exact solutions are
rare, and cannot be relied upon for a general theory. Various
perturbative approaches to quantum critical dynamics have been
developed\cite{ssrelax,damle}, and these usually rely upon
effective classical models of order-parameter or quasiparticle
dynamics as a point of departure.

We will begin our discussion in Section~\ref{sec:ising} by
considering the quantum Ising chain. An essentially exact
description of the long time dynamics is available\cite{apy} in
all the different low temperature regions of the phase diagram of
this model, and this will allow us to describe and distinguish the
essential characteristics of the order-parameter, quasiparticle,
and quantum-critical dynamics. Section~\ref{sec:did} will consider
a quantum critical point in two dimensions between two
superconductors with distinct pairing symmetries. The critical
theory involves interacting fermionic and bosonic excitations, and
we will discuss approximate theories for the quantum critical
dynamics and their possible relevance to recent photo-emission
experiments on the cuprate superconductors. Finally, in
Section~\ref{sec:sy} we will discuss the quantum critical dynamics
in simplified models of correlated electrons in the presence of
quenched disorder, and their relationship to experiments on the
heavy fermion superconductors.

The reader is also referred to another recent
review\cite{altenberg} which covers additional topics on quantum
phase transitions presented in my talk at MB-11.

\section{Quantum Ising chain}
\label{sec:ising}

\begin{figure}
\epsfxsize=4in \centerline{\epsfbox{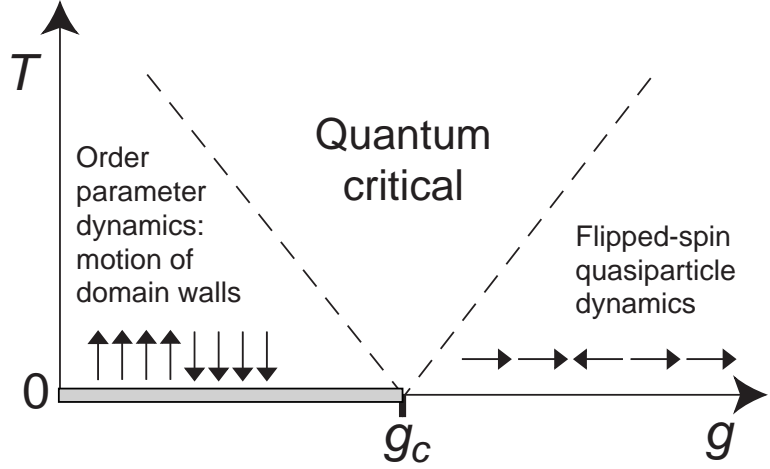}} \caption{Phase
diagram of the quantum Ising chain $H_I$ in (\protect\ref{hi}).
There is long range order, with $N_0 \neq 0$ in
(\protect\ref{lro}), only for $g<g_c$ and $T=0$ on the shaded
line. The dashed lines represent crossovers at $\Delta \sim k_B
T$. Quasiclassical models for the long times dynamics in the lower
left (Section~\protect\ref{sec:op}) and lower right
(Section~\protect\ref{sec:qp}) regions are available. The quantum
critical dynamics is discussed in Section~\protect\ref{sec:qc}.}
\label{fig1}
\end{figure}

We consider one of the simplest models which displays a
second-order quantum phase transitions; its phase diagram is
nevertheless rich enough to display all the dynamic regimes we
wish to distinguish. The Hamiltonian of the quantum Ising chain is
\begin{equation}
H_I = -\sum_j \left( \sum_{\ell >0} J_{\ell} \sigma^z_j
\sigma^z_{j+\ell} + g \sigma^x_j \right) \label{hi}
\end{equation}
where $\sigma^z_j$, $\sigma^x_j$ are Pauli matrices on a chain of
sites $j$, $J_{\ell}$ $(> 0)$ are short-ranged exchange constants,
and $g$ $(>0)$ is the transverse field. The operator $\prod_j
\sigma^x_j$ commutes with $H_I$, and generates its global $Z_2$
Ising symmetry. We will consider the phase diagram\cite{apy} of
$H_I$ as a function of $g$ and $T$, shown in Fig~\ref{fig1}.
The global $Z_2$ symmetry is spontaneously broken at $T=0$ for $g
< g_c$, where $g_c$ is a critical value of order the $J_{\ell}$.
There is a second order phase transition at $g=g_c$, $T=0$, and
the symmetry is restored elsewhere in the phase diagram. We
describe the different regimes of Fig~\ref{fig1} in the following
subsections.

\subsection{Order parameter dynamics}
\label{sec:op}

First consider $H_I$ at $g=0$. Here $H_I$ is the classical Ising
model and all states are known exactly, as is the $T>0$ partition
function. The ground states are doubly degenerate, $\prod_j
|\uparrow\rangle_j$ or $\prod_j |\downarrow \rangle_j$, and have
long-range ferromagnetic order in $\sigma^z_j$. All excited states
can also be written down exactly, and the lowest energy ones are
domain walls between the two ordered states
\begin{equation}
\ldots |\uparrow \rangle_{j-2} |\uparrow \rangle_{j-1}|\uparrow
\rangle_{j}|\downarrow \rangle_{j+1}|\downarrow
\rangle_{j+2}|\downarrow \rangle_{j+3} \dots, \label{domain}
\end{equation}
whose energy is $\Delta=2\sum_{\ell >0} \ell J_{\ell}$ above the
ground state. The entire set of excited states can be built out of
multiple domain wall states. At a small $T>0$, there is an
exponentially small density $\rho \sim e^{-\Delta/k_B T}$ of
thermally excited domain walls. A simple calculation shows that
the presence of these excitations is sufficient to destroy the
long range order beyond an exponentially large correlation length
$\xi = 1/2\rho$.

Now move to a small $g>0$. All the states are now perturbed by
quantum fluctuations, but the qualitative picture remains the same
at long length scales. There is still ferromagnetic order in the
ground state, and a broken $Z_2$ symmetry, with the long range
correlation
\begin{equation}
\lim_{|j-k| \rightarrow \infty} \langle \sigma^z_j \sigma^z_k
\rangle_{T=0} = N_0^2 ,\label{lro}
\end{equation}
but $N_0$ is no longer unity as it is in the classical model;
instead quantum fluctuations reduce $N_0$ with increasing $g$
until it ultimately vanishes at $g=g_c$. This subsection will
restrict attention to $g<g_c$. The degenerate manifold of the
domain wall states (\ref{domain}) now broaden into a band of
single particle states with dispersion $\varepsilon_p = \Delta +
4g \sin^2 (p/2) + {\cal O} (g^2) $ as function of momentum $p$; we
assume unit lattice spacing) and have now redefined $\Delta$ to be
the true energy gap including all corrections. The resulting
motion and collision of the thermally excited domain walls is
responsible for the long time relaxational dynamics of the
ferromagnetic order parameter. This picture was used to develop a
simple solvable quasiclassical theory\cite{apy} for the dynamic
correlations of the order parameter in the $g<g_c$, low $T$ region
of Fig~\ref{fig1}, and the results are most conveniently expressed
in terms of the dynamic structure factor $S(p, \omega)$:
\begin{equation}
S(p,\omega) = \sum_j \int_{-\infty}^{\infty} dt e^{-i(pj-\omega
t)} \mbox{Tr} \left. \left( e^{-H_I /k_B T} e^{i H_I t/\hbar}
\sigma^z_j e^{-i H_I t/\hbar} \sigma^z_0 \right)\right/ \mbox{Tr}
\left(e^{-H_I /k_B T} \right) \label{defS}
\end{equation}
At $T=0$, the long-range order (\ref{lro}) leads to the elastic
Bragg peak
\begin{equation}
S(p,\omega) = (2\pi)^2 N_0^2 \delta(p) \delta(\omega) + \ldots
\label{delta}
\end{equation}
where the ellipsis represents contributions at frequences $\omega
> 2\Delta$ associated with the creation of two or more domain
walls by the external probe. At $T>0$ the quasiclassical order
parameter dynamics broadens the delta functions in (\ref{delta}),
and leads to
\begin{equation}
S(p,\omega) =  N_0^2 \int_{-\infty}^{\infty} dx
\int_{-\infty}^{\infty} dt e^{-i(px-\omega t)} R(x,t) + \ldots
\label{delta2}
\end{equation}
where $R(x,t)$ is a relaxational function. The form of $R(x,t)$
for general $x$, $t$ is involved, but its basic features can be
understood by looking at two simple limits:
\begin{eqnarray}
R(x,0) = e^{-|x|/\xi}~~~&;&~~~\frac{1}{\xi} = 2 \int_{-\pi}^{\pi}
\frac{dp}{2 \pi} e^{-\varepsilon_p/k_B
T}\nonumber \\
R(0, t) = e^{-|t|/\tau_{\varphi}}~~~&;&~~~\frac{1}{\tau_{\varphi}}
= \frac{2 k_B T}{\pi \hbar} e^{-\Delta/k_B T}. \label{rxt}
\end{eqnarray}
So $R$ decays exponentially on a characteristic spatial scale
$\xi$, and a characteristic temporal scale, the coherence time
$\tau_{\varphi}$; the delta functions in (\ref{delta}) therefore
broaden at $T>0$ to a momentum width of order $1/\xi$, and a
frequency width of order $1/\tau_{\varphi}$. The expression
(\ref{rxt}) contains an exact result for $\tau_{\varphi}$ at low
$T$: the quantum Ising chain, and the closely related dilute Bose
gas\cite{korepin,apy}, are the only interacting many body quantum
systems for which exact results for such a relaxation rate are
available. Remarkably, $\tau_{\varphi}$ involves only the energy
gap $\Delta$ and fundamental constants of nature; it also
satisfies $\tau_{\varphi} \gg \hbar/k_B T$ as is required for the
applicability of the quasiclassical theory.

\subsection{Quasiparticle dynamics}
\label{sec:qp}

Next consider the opposite limit of very large $g$. At $g=\infty$
all eigenstates can again be written down simply. There is a
unique, paramagnetic ground state $\prod_{j} |\rightarrow
\rangle_j$, where $|\rightarrow \rangle_j = (|\uparrow \rangle_j +
|\downarrow \rangle_j)/\sqrt{2}$ is an eigenstate of $\sigma^x_j$.
In strong contrast to the states at $g=0$, equal time correlations
of $\sigma^z_j$ are now non-zero only on-site, and so the spin
correlation length is effectively zero. The lowest excited states
are obtained by flipping a single spin:
\begin{equation}
\ldots |\rightarrow \rangle_{j-2} |\rightarrow
\rangle_{j-1}|\leftarrow \rangle_{j}|\rightarrow
\rangle_{j+1}|\rightarrow \rangle_{j+2}|\rightarrow \rangle_{j+3}
\dots, \label{flip}
\end{equation}
where $|\leftarrow \rangle_j = (|\uparrow \rangle_j - |\downarrow
\rangle_j)/\sqrt{2}$ is the other eigenstate of $\sigma^x_j$, and
these states have energy $2g$ above the ground state. We can now
examine corrections to the limit case in powers of $1/g$. The
ground state remains paramagnetic but the spin correlation length
increases as $g$ is decreased. The flipped spin states
(\ref{flip}) again broaden into a quasiparticle band with
$\varepsilon_p = \Delta + 4\sum_{\ell>0} J_{\ell} \sin^2 (p
\ell/2) + {\cal O} (g^2)$ and $\Delta = 2g -
2\sum_{\ell>0}J_{\ell}+ {\cal O} (g^2)$. Keep in mind, however,
that the physical interpretation of these quasiparticles is
entirely distinct from those of Section~\ref{sec:op}: they are not
domain walls, but localized spin flips. A crucial property of the
quasiparticles is that they have an infinite lifetime for a finite
range of momenta around $p=0$: this is simply because energy and
momentum conservation prohibit their decay into any other states.
Consequently, although equal time correlations of $\sigma^z_j$
decay very rapidly in space, there are long-range correlations in
spacetime; this is evident from the expression for the dynamic
structure factor
\begin{equation}
S(p, \omega) = \pi {\cal A} \delta(\omega - \varepsilon_p/\hbar) +
\ldots~~~;~~~T=0, \label{pole}
\end{equation}
associated with quasiparticle pole, as infinite range temporal and
spatial correlations are needed for the Fourier transform to
conspire to have a pole. Here the ellipsis represents
multiparticle contributions at frequencies $\omega > 3 \Delta$.
The quasi-particle residue ${\cal A}$ is finite in the
paramagnetic phase and decreases as $g$ is decreased.

At $T>0$, there will again be an exponentially small density of
these thermally excited quasiparticles. A quasiclassical transport
equation for the collisions of these quasiparticles can be written
down and, remarkably, solved exactly. This solution shows that the
quasiparticle collisions broaden (\ref{pole}) into a Lorentzian
\begin{equation}
S(p, \omega) = \frac{{\cal A}/\tau_{\varphi}}{(\omega -
\varepsilon_p/\hbar)^2 + (1/\tau_{\varphi})^2}.
\end{equation}
The expression for the quasiparticle-width, $1/\tau_{\varphi}$,
turns out to be identical to that in (\ref{rxt}), where $\Delta$
is now the energy gap to the flipped-spin quasiparticles. Note
that the characteristic collision time is again larger than
$\hbar/k_B T$, as is needed to justify a classical dynamical
model.

\subsection{Quantum critical dynamics} \label{sec:qc}

The quasiclassical models in Sections~\ref{sec:op}
and~\ref{sec:qp} provide an essentially exact description of the
long time dynamics as $T \rightarrow 0$ for any fixed $g$, other
than at the critical point $g=g_c$. As illustrated in
Fig~\ref{fig1}, this failure broadens into a wide quantum critical
region at larger $T$, which is, in principle, easily accessible in
experiments.

We begin our discussion of spin correlations in this region by
first considering the single point $g=g_c$, $T=0$. The energy gap
vanishes here as $\Delta \sim |g-g_c|$. Also as $g \nearrow g_c$,
the ferromagnetic order parameter vanishes\cite{pfeuty} as $N_0
\sim (g_c - g)^{1/8}$, and so it is not sensible anymore to think
in terms of domain walls. Similarly, for $g \searrow g_c$, the
quasiparticle residue vanishes\cite{apy} as ${\cal A} \sim
(g-g_c)^{1/4}$, and so the flipped spin quasiparticles are also
not well defined at $g=g_c$. A less intuitive description of the
long time dynamics must be developed here. The fundamental
property of the critical point that enables a description of the
dynamics is that of scale and conformal invariance. One signal of
this invariance is the long distance behavior of the equal time
$\sigma^z$ correlation
\begin{equation}
\langle \sigma^z_j \sigma^z_k \rangle =
\frac{Z}{|j-k|^{1/4}}~~~;~~~g=g_c,~T=0, \label{power}
\end{equation}
where $Z$ is a non-zero, non-universal number. Remarkably, the
powerful technology of conformal invariance allows one to
reconstruct the exact long distance and long time spin correlation
function at $T>0$ using as input only the form (\ref{power}) (such
`wizardry' is unfortunately not possible in higher dimensions).
The answer is more easily expressed in terms of the dynamic
susceptibility, $\chi (p, \omega)$, which is related to
$S(p,\omega)$ in (\ref{defS}) by the usual fluctuation dissipation
theorem:
\begin{equation}
\chi (p, \omega) = \left( \frac{\Gamma(7/8) c^{3/4}}{2^{7/4}
\pi^{3/4} \Gamma (1/8)} \right) \frac{Z}{(k_B T/\hbar)^{7/4}}
\frac{ \Gamma \left( \displaystyle \frac{1}{16} -i
\frac{\omega+cp}{4 \pi k_B T/\hbar} \right)\Gamma \left(
\displaystyle \frac{1}{16} -i \frac{\omega-cp}{4 \pi k_B T/\hbar}
\right)}{\Gamma \left( \displaystyle \frac{15}{16}-i
\frac{\omega+cp}{4 \pi k_B T/\hbar} \right)\Gamma \left(
\displaystyle \frac{15}{16}-i \frac{\omega-cp}{4 \pi k_B T/\hbar}
\right)} \label{conformal}
\end{equation}
where $c$ is the non-universal velocity of excitations at the
critical point. This is a rather cumbersome expression whose
functional form is not terribly intuitive. Its structure is
nevertheless rather simple, and this becomes clear from the
following expression which provides an excellent fit to
(\ref{conformal}) over a wide window of low frequencies or
wavevectors (this is also the region over which the spectral
density $\mbox{Im} \chi(p,\omega)/\omega$ has significant weight):
\begin{equation}
\chi(p,\omega) = \frac{\chi(0,0)}{1 - i (\omega/\Gamma_R) + p^2
\xi^2 - (\omega/\omega_1)^2}. \label{fit}
\end{equation}
The relaxation rate $\Gamma_R$ is given by
\begin{eqnarray}
\Gamma_R &\equiv& \left( i \chi(0,0) \left. \frac{\partial
\chi^{-1} (0,\omega)}{\partial \omega},
\right|_{\omega=0}\right)^{-1} \nonumber \\
&=& \left( 2 \tan \frac{\pi}{16} \right) \frac{k_B T}{\hbar}
\label{gr}
\end{eqnarray}
where we used (\ref{conformal}) in the first equation. As claimed
in the introduction, this characteristic frequency is a universal
number times $k_B T/\hbar$. We determined $\omega_1$ and $\xi$ by
fitting (\ref{fit}) to (\ref{conformal}) over the range $0 <
\omega,cp < 2 k_B T/\hbar$ (the fit was essentially perfect) and
obtained the best fit values
\begin{equation}
\omega_1 = 0.795 k_B T/\hbar~~~;~~~\xi = 1.280 \hbar c/k_B T.
\end{equation}
As expected, all time and length scales are determined by $T$
alone. Notice that (\ref{fit}) has the structure of the response
function of the conventional van Hove dynamics of an overdamped
field\cite{forster}. What is novel here the characteristic damping
frequency is not an adjustable phenomenological parameter
depending upon the coupling to some heat bath, but a universal
rate determined by the absolute temperature and fundamental
constants of nature.

At large frequencies, $\hbar |\omega| \gg k_B T$, and wavevectors,
$\hbar c |p| \gg k_B T$, the expression (\ref{conformal}) reduces
to
\begin{equation}
\chi (p, \omega) = \frac{\Gamma(7/8) 2^{7/4} \pi
c^{3/4}}{\Gamma(1/8)} \frac{Z}{(c^2 p^2 - \omega^2)^{7/8}}
\label{bci}
\end{equation}
This represents the spectrum at the gapless critical point
$g=g_c$, $T=0$; as expected, the complex frequency plane has no
quasiparticle pole, but a branch-cut originating at $\omega = \pm
cp$ associated with the continuum of critical excitations. For
practical purposes, it is worth noting that the spectral weight in
this critical region is rather small, and experimental
observations at $T>0$ will be dominated by the relaxation spectrum
in (\ref{fit}).

\section{Transition between BCS superconductors}
\label{sec:did}

This section reviews quantum critical dynamics in a
two-dimensional model which may be of relevance to the cuprate
high temperature superconductors. Observation\cite{valla} of
quantum critical behavior in the spectral function of the
fermionic nodal quasiparticles stimulated our study
\cite{vojta1,vojta2} of a wide class of possible models which
could yield a suitable quantum critical point. It turned out to be
rather difficult to satisfy the needed requirements: a quantum
critical point below its upper critical dimension, with nodal
quasiparticles a primary degree of freedom. Only two such
candidates were found, both involving transitions with changes in
the pairing symmetry of BCS superconductors in two dimensions: the
transitions from a $d_{x^2-y^2}$ superconductor to either a
$d_{x^2-y^2}+id_{xy}$ or a $d_{x^2-y^2}+is$ superconductor. We
considered the first of these more likely on microscopic grounds
and will review its theory here; the theory for the second is
closely related. We also note that our interpretation of the
spectra of the cuprates does not require that such a quantum phase
transition be actually present on the physical axis as a function
of doping; all that is required is that a quantum critical point
be nearby in the generalized parameter space.

Recent tunnelling measurements\cite{dagan} have provided support
for a transition from a $d_{x^2-y^2}$ superconductor at optimal
doping to a $d_{x^2-y^2}+id_{xy}$ superconductor in the overdoped
regime, at least in thin films of
Y$_{1-y}$Ca$_y$Ba$_2$Cu$_3$O$_{7-x}$. This sequence of transitions
may be understood on the basis of some early theoretical work.
While there is strong theoretical evidence for $d_{x^2-y^2}$
superconductivity in lightly doped antiferromagnets\cite{metzner},
studies of the Hubbard model in limit of very large hole
density\cite{maxim1,maxim2} (low electron density) showed instead
an instability to $d_{xy}$ superconductivity. Interpolating
between these limits, we can expect\cite{altenberg} that a
$d_{x^2-y^2}+id_{xy}$ superconductor appears an intermediate
phase.

\begin{figure}
\epsfxsize=2.5in \centerline{\epsfbox{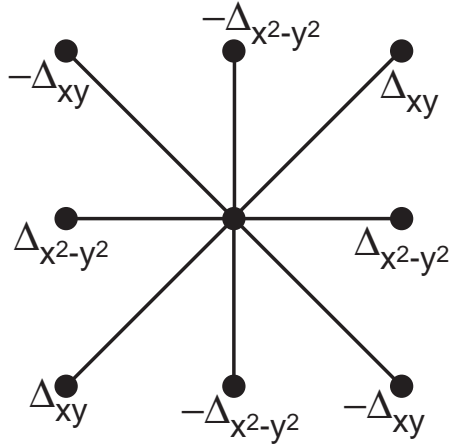}} \caption{Values
of the pairing amplitudes, $-\langle c_{i \uparrow} c_{j
\downarrow} -  c_{i \downarrow} c_{j \uparrow} \rangle$ with $i$
the central site, and $j$ is one of its 8 near neighbors.}
\label{fig2}
\end{figure}

As we are interested only in universal critical properties, we can
develop a theory using the simplest phenomenological model which
displays the required transition. So we consider the Hamiltonian
\begin{equation}
{H}_{tJ} = \sum_{k} \varepsilon_k c_{k \sigma}^{\dagger} c_{k
\sigma} + J_1 \sum_{\langle ij \rangle} {\bf S}_i \cdot {\bf
S}_{j} + J_2 \sum_{{\rm nnn}~ij} {\bf S}_i \cdot {\bf S}_j,
\label{g8}
\end{equation}
where $c_{j\sigma}$ is the annihilation operator for an electron
on the square lattice site $j$ with spin
$\sigma=\uparrow,\downarrow$, $c_{k\sigma}$ is its Fourier
transform to momentum space, $\varepsilon_k$ is the dispersion of
the electrons,
\begin{equation}
S_{j\alpha} = \frac{1}{2} c^{\dagger}_{j\sigma}
\sigma_{\sigma\sigma'}^{\alpha} c_{j \sigma'} \label{g2a}
\end{equation}
with $\sigma^{\alpha}$ the Pauli matrices, and $J_1$, $J_{2}$ are
first and second neighbor antiferromagnetic exchange interactions
which induce the superconductivity. We apply the standard BCS
theory to $H_{tJ}$: this will yield an adequate description of the
low temperature properties, except in the vicinity of the quantum
critical point. The BCS Hamiltonian is
\begin{eqnarray}
{H}_{BCS} = \sum_{k} \varepsilon_k c_{k \sigma}^{\dagger} c_{k
\sigma} &-& \frac{J_1}{2} \sum_{j,\mu} \Delta_{\mu}
(c_{j\uparrow}^{\dagger} c_{j+\hat{\mu},\downarrow}^{\dagger} -
c_{j\downarrow}^{\dagger} c_{j+\hat{\mu},\uparrow}^{\dagger}) +
\mbox{h.c.} \nonumber \\ &-& \frac{J_2}{2} {\sum_{j,\nu}}^{\prime}
\Delta_{\nu} (c_{j\uparrow}^{\dagger}
c_{j+\hat{\nu},\downarrow}^{\dagger} - c_{j\downarrow}^{\dagger}
c_{j+\hat{\nu},\uparrow}^{\dagger}) + \mbox{h.c.}, \label{g9}
\end{eqnarray}
where the first summation over $\mu$ is along the nearest neighbor
directions $\hat{x}$ and $\hat{y}$, while the second summation
over $\nu$ is along the diagonal neighbors $\hat{x}+\hat{y}$ and
$-\hat{x}+\hat{y}$. To obtain $d_{x^2-y^2}$ and $d_{xy}$ pairing,
we choose $\Delta_x = - \Delta_y = \Delta_{x^2-y^2}$, and
$\Delta_{x+y} = - \Delta_{-x+y} = \Delta_{xy}$. We summarize our
choices for the spatial structure of the pairing amplitudes (which
determine the Cooper pair wavefunction) in Fig~\ref{fig2}.
The values of $\Delta_{x^2-y^2}$ and $\Delta_{xy}$ are to be
determined by minimizing the ground state energy:
\begin{equation}
E_{BCS} = J_1 |\Delta_{x^2-y^2}|^2 +J_2 |\Delta_{xy}|^2 - \int
\frac{d^2 k}{4 \pi^2} \left[ \sqrt{\varepsilon_k^2 + |\Delta_k|^2}
- \varepsilon_k \right] \label{g10}
\end{equation}
where the pairing amplitude in momentum space is
\begin{equation}
\Delta_k = J_1 \Delta_{x^2-y^2}(\cos k_x - \cos k_y) + 2 J_2
\Delta_{xy} \sin k_x \sin k_y . \label{g11}
\end{equation}
Notice that the energy depends upon the relative phase of
$\Delta_{x^2-y^2}$ and $\Delta_{xy}$: this phase is therefore an
observable property of the ground state.

The minimization of $E_{BCS}$ was carried out in
Ref.~\citelow{altenberg}. For small $J_2/J_1$ the ground state was
a $d_{x^2-y^2}$ superconductor with $\Delta_{x^2-y^2}\neq 0$ and
$\Delta_{xy}=0$. Above a critical value of $J_2/J_1$ there was a
continuous transition to a state with $\Delta_{x^2-y^2}\neq 0$,
$\Delta_{xy} \neq 0$, and
\begin{equation}
\arg (\Delta_{xy}) = \arg (\Delta_{x^2-y^2}) \pm \pi/2.
\label{g12}
\end{equation}
This is a $d_{x^2-y^2}+id_{xy}$ superconductor. The choice of the
sign in (\ref{g12}) is associated with the breaking of time
reversal symmetry, and this suggests that the transition can be
described by a $Z_2$ Ising order parameter. As usual, we can
associate the Ising order with a real, scalar field $\phi$ (the
coarse-grained value of the local Ising order) which we can
identify here by $\phi = i \Delta_{xy}$, in the gauge where
$\Delta_{x^2-y^2}$ is real. However, the critical theory of the
transition is {\em not} simply that of the quantum Ising model in
2+1 dimensions {\em i.e.} not that of the two-dimensional version
of the model considered in Section~\ref{sec:ising}. The fermionic
quasi-particles are also primary critical degrees of freedom, and
the required quantum field theory couples the Ising field $\phi$
to fermionic fields representing the lowest energy Bogoliubov
quasiparticles. We will not explicitly write down this field
theory here, and instead refer the reader to another recent
review\cite{altenberg} by the author.

\begin{figure}
\epsfxsize=4.5in \centerline{\epsfbox{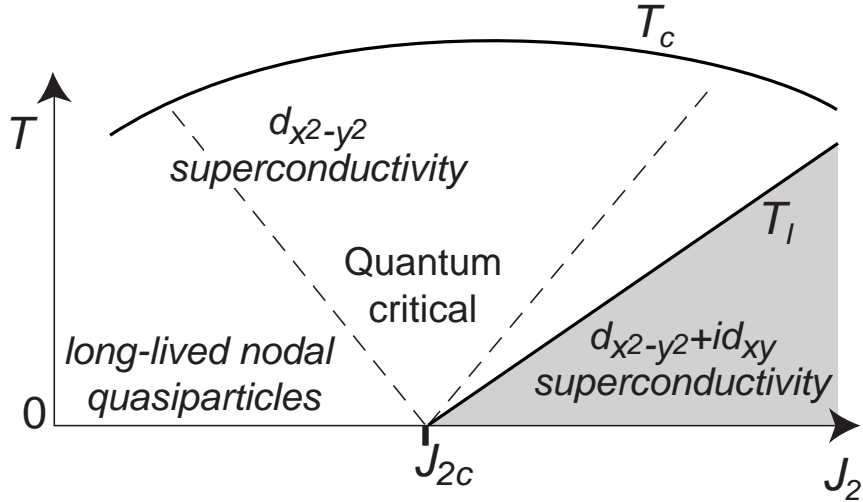}} \caption{Phase
diagram of $H_{tJ}$. As in Fig~\protect\ref{fig1}, dashed lines
represent crossovers, but the full lines are true thermodynamic
phase transitions. There is a quantum critical point at $T=0$,
$J_2= J_{2c}$, but now it extends into a line of second order
phase transitions at temperatures $T_I > 0$. The critical theory
for the $T=0$ quantum critical point was reviewed in
Ref.~\protect\citelow{altenberg} and involves an Ising field and
fermionic degrees of freedom. The $T_I$ line is in the
universality class of the classical two-dimensional Ising model.
The ground state is superconducting everywhere below $T_c$, with
$d_{x^2-y^2}+id_{xy}$ superconductivity in the shaded region, and
$d_{x^2-y^2}$ superconductivity elsewhere. The crossover
boundaries and $T_I$ approach the $T=0$, $J_2 = J_{2c}$ critical
point linearly because the best estimate\protect\cite{kvesch,qmc}
for the exponent $z\nu$ is 1.00. } \label{fig3}
\end{figure}

The finite temperature phase diagram of $H_{tJ}$ in the vicinity
of the critical point is sketched in Fig~\ref{fig3}.
We discuss below the nature of the fermion Green's function in the
different regimes below the superconducting critical temperature
$T_c$.

We can write the fermion Green's function in the superconductor
compactly using the Nambu notation. We define the spinor field
$\Psi_k = (c_{k\uparrow}, c_{-k\downarrow}^{\dagger})$. Then the
retarded $\Psi_k$ Green's function $G(k, \omega)$ can be written
as
\begin{equation}
\hbar G^{-1} (k, \omega) = \hbar\omega - \varepsilon_k \tau^z +
\mbox{Re}(\Delta_k) \tau^x - \mbox{Im} (\Delta_k) \tau^y -
\Sigma(k, \omega) \label{gs}
\end{equation}
where $\tau^{x,y,z}$ are Pauli matrices in Nambu particle-hole
space. The above expression follows from the BCS theory
(\ref{g9}), with the self energy $\Sigma$ representing the effects
of interaction between the quasiparticles and any additional
collective modes.

Consider first the low $T$ region with $J_2 < J_{2c}$ in
Fig~\ref{fig3}. Here the ground state is well described by the BCS
theory of the $d_{x^2-y^2}$-wave superconductor. We can compute
the effects of collisions between the fermionic quasiparticles in
very much the same spirit as in Landau's Fermi liquid
theory\cite{landau}: the main difference is that instead of a
whole Fermi surface, we now have isolated `nodal' Fermi points at
four points in the Brillouin zone determined by the solution of
$\varepsilon_k=0, \Delta_k=0$. Consequently, the density of states
for low-energy scattering of quasiparticles is even smaller than
in a Fermi liquid, and the quasiparticles remain well-defined
excitations. Right at the Fermi level, and at low $T$, these
collisions lead to $\Sigma \sim i T^3$, which is smaller than the
result $\sim i T^2$ in a conventional Fermi liquid. A
quasiclassical description of the fermionic quasiparticles
therefore remains valid, in the same framework used in Landau's
Fermi liquid theory.

Next, we move to the opposite low $T$ region in Fig~\ref{fig3}
with $J_2 > J_{2c}$ and $d_{x^2-y^2}+id_{xy}$ superconductivity.
Here time-reversal symmetry is broken with $\langle \phi \rangle
\neq 0$. The theory of the quasiparticles is very similar to that
in the small $J_2$ region above with one very important
difference. The equations $\varepsilon_k=0, \Delta_k=0$ now have
no solution for any $k$ and so the nodal points have been gapped
out. There are still low-energy fermionic excitations near the
position of the original nodal points, but these appear only above
an energy gap $\Delta$. As in Section~\ref{sec:ising}, this energy
gap vanishes as we approach the critical point as $\Delta \sim
(J_2 - J_{2c})^{z\nu}$, and the best estimate\cite{kvesch,qmc} for
the critical exponent is $z\nu \approx 1.00$. As long as $T
\rightarrow 0$ at any fixed $\Delta > 0$, the density of thermally
excited quasiparticles is exponentially small, and so $\Sigma \sim
e^{-\Delta/T}$. So again a quasiclassical transport theory of the
quasiparticle collisions applies.

Finally, we turn to the intermediate temperature quantum critical
region in Fig~\ref{fig3}, in the vicinity of the $J_{2}= J_{2c}$
critical point. Here the physics is rather similar to the
corresponding region in Section~\ref{sec:qc}, but we no longer
have the benefit of an exact solution for the spectral function.
Strong fluctuations of the Ising order parameter $\phi$, whose
condensate gapped the nodal points in the $J_2 > J_{2c}$ region,
now lead to strongly damped fermionic excitations. Loosely
speaking, rapid temporal oscillations between
$d_{x^2-y^2}+id_{xy}$ and $d_{x^2-y^2}-id_{xy}$ pairing lead to
broad spectral functions. A theory for the low frequency form of
the fermion spectral function was developed in
Ref.~\citelow{vojta1}, using an expansion in $(3-d)$ (where $d$ is
the spatial dimensionality) and phenomenolgical ansatzes similar
to (\ref{fit}) for the frequency and wavevector dependence of the
spectral function of the critical excitations. At low frequencies
near the Fermi level, the result can be written in the form
(compare (\ref{conformal}) and (\ref{fit}))
\begin{equation}
G^{-1} (k, \omega) = Z^{-1} T^{\eta_f} \left( \hbar \omega -
\varepsilon_k \tau^z + \Delta_k \tau^x + i \hbar \Gamma_R \right),
\label{gs1}
\end{equation}
where $\eta_f$ is a critical exponent, $Z$ is a non-universal,
non-singular prefactor, and we can take $\Delta_k$ real because
$\langle \phi \rangle =0$ in the quantum critical region. The
estimates for the exponent $\eta_f$ are $\eta_f \approx (3-d)/14$
in the $(3-d)$ expansion\cite{vojta1}, and  $\eta_f \approx 1/(3
\pi^2 N)$, with $N=2$, in the $1/N$ expansion\cite{kvesch}. The
damping rate $\Gamma_R$ has the same remarkable universal
structure we found in Section~\ref{sec:qc}. Just as in (\ref{gr}),
we now find\cite{vojta1,kvesch} that $\Gamma_R$ is universally
related to the absolute temperature, with
\begin{equation}
\Gamma_R = 0.581 \frac{k_B T}{\hbar}
\end{equation}
in the $(3-d)$ expansion.

It is important to keep in mind that (\ref{gs1}) is a low
frequency form and does not hold at frequencies which are much
larger than $k_B T/\hbar$. In the latter regime we obtain a result
which is similar in structure to that obtained by taking the large
$\omega$ limit of the exact result (\ref{conformal}) of the
quantum Ising chain. Near one of the nodal points we find
\begin{equation}
G(p, \omega) = Z C_f \frac{-\omega - c p_x \tau^z - c p_y
\tau^x}{( c^2 p^2 - \omega^2 )^{1 - \eta_f /2}} \label{bc}
\end{equation}
where $p$ measures momentum deviation from one of the nodal points
and the $p$ co-ordinates have been rotated by 45 degrees from the
$k$ co-ordinates, $C_f$ is a computable universal number, and $c$
is a velocity. In principle the velocities appear before the
$\tau^z$ and $\tau^x$ should be different, but they become equal
in the asymptotic region near the critical point \cite{vojta1}.
The spectrum (\ref{bc}) is the analog of (\ref{bci}), and its most
important property is, of course, that it does not have
quasi-particle pole, but only a branch cut representing the
continuum of critical excitations.

\section{Disordered spin systems}
\label{sec:sy}

The plethora of intermetallic ``heavy fermion'' compounds have
provided a fertile ground for the study for the study of quantum
phase transitions and the associated quantum critical region at
finite temperature. Useful reviews of the current experimental and
theoretical status have been provided recently by Coleman and
collaborators\cite{piers}. Most of these systems are near a zero
temperature magnetic ordering transition of some type, and this
has allowed investigation of quantum criticality using a variety
of probes. Among the most thoroughly investigated compounds is
CeCu$_{6-x}$Au$_x$: neutron scattering and thermodynamic
measurements\cite{almut} show convincing signs of universal
critical behavior, with a characteristic frequency scale of order
$k_B T/\hbar$ and scaling of dynamic response functions as a
function of $\hbar\omega/k_B T$. This behavior is not
compatible\cite{book,piers} with spin-density wave
theories\cite{hertz,millis}, which consider perturbative
corrections from interaction between collective paramagnon modes,
and the search for the appropriate quantum-critical model is an
active topic of current research. An important open question in
these investigations is the importance of quenched disorder to the
observed critical behavior. Theories for the onset of spin glass
order in metallic systems have been proposed (see
Ref.~\citelow{royal} for a review), and glassy dynamics has been
observed\cite{maclaughlin1} in UCu$_{5-x}$Pd$_{x}$, a compound
which also displays $\hbar \omega/k_B T$ scaling in neutron
scattering\cite{aronson}. More unexpectedly, the glassy behavior
persists at compositions which are nominally
stoichiometric\cite{maclaughlin2} ($x=1$).

Here we shall review a class of theories which have their origin
in studies of systems with strong quenched disorder. Analyses of
disordered models of frustrated spin systems\cite{sy2,gps} and
doped Mott insulators\cite{gp} have shown that a useful starting
point for the critical theory is a seemingly simple model which we
shall describe below. It consists  a single quantum spin
interacting with a `bath' of low energy collective spin
excitations which represent a mean-field description of its
environment. Generalized models have recently been argued to apply
to systems in two dimensions even in the absence of quenched
disorder\cite{si}.

We consider the following single spin quantum partition function,
which defines a well-posed, quantum mechanical problem in its own
right:
\begin{eqnarray}
{\cal Z} = \int {\cal D} {\bf n} (\tau) \delta({\bf n}^2 (\tau) -
1) && \exp \left( - i S \int_0^{\beta}  A_{\tau} ({\bf n}(\tau)) d
\tau \right. \nonumber \\
&&~~~\left.+ \int_0^{\beta} d \tau \int_{0}^{\beta} d
\tau^{\prime} D(\tau - \tau^{\prime}) {\bf n}(\tau) \cdot {\bf n}
(\tau^{\prime}) \right).
\end{eqnarray}
Here $\tau$ is imaginary time which runs periodically from 0 to
$\beta = \hbar/k_B T$, ${\bf n} (\tau)$ is the orientation of the
quantum spin with ${\bf n} (\tau+\beta) = {\bf n} (\tau)$, and $S$
its spin quantum number with $2S =$ integer. The first term in
${\cal Z}$ is the Berry phase of the spin, with $A_{\tau} ({\bf n}
(\tau)) d\tau$ equal to the area of the spherical triangle on the
unit sphere with vertices at ${\bf n} (\tau)$, ${\bf n} (\tau +
d\tau)$, and a fixed (but arbitrary) reference direction ${\bf
n}_0$. The function $D(\tau)$ (also periodic with period $\beta$)
is the retarded self-interaction of the spin and represents the
influence of the spin environment. The model ${\cal Z}$ was
introduced and examined in a large $N$ expansion\cite{sy2}, and
useful renormalization group analyses were
presented\cite{anirvan,qmsi}. In its application as a mean-field
theory of bulk lattice models, the solution of ${\cal Z}$ usually
has to be supplemented by a self-consistency condition for
$D(\tau)$: the simplest of these is $D(\tau-\tau^{\prime}) =
(J/\hbar)^2 \langle {\bf n} (\tau) \cdot {\bf n} (\tau')
\rangle_{{\cal Z}}$, where $J$ is an energy scale for the exchange
interactions, but more complicated self-consistency conditions
have also been considered\cite{gp,gps,si}.

In a quantum critical system, we may expect power-law correlations
of all observables at $T=0$. So it is of particular interest to
examine ${\cal Z}$ for the case where
\begin{equation}
D(\tau) \sim \frac{1}{|\tau|^{2-\alpha}} \label{da}
\end{equation}
for large $|\tau|$ at $T=0$. Non-trivial spin correlations emerge
for $\alpha > 0$, and the renormalization group
analysis\cite{anirvan,qmsi} proceeds in powers of $\alpha$.
Although this is an interacting quantum field theory and it is not
possible to compute the flow equations exactly, it was recently
shown\cite{vbs,gps} that the structure of the quantum theory
allowed one to deduce some results to all orders in $\alpha$; in
particular it was shown that for the case where $D(\tau )$ obeyed
(\ref{da}),
\begin{equation}
\langle {\bf n} (\tau) \cdot {\bf n} (\tau') \rangle_{{\cal Z}}
\sim \frac{1}{|\tau-\tau^{\prime}|^{\alpha}}. \label{nn}
\end{equation}
This is a non-trivial result\cite{sy2,anirvan,qmsi,vbs,gps}, and
depends crucially on the the Berry phase term in ${\cal Z}$.
Comparison of (\ref{nn}) and (\ref{da}) with the particular
self-consistency condition mentioned above selects the value
$\alpha=1$, and this is the value that has appeared frequently in
physical applications\cite{sy2,gp,gps,si}.

\begin{figure}
\epsfxsize=5in \centerline{\epsfbox{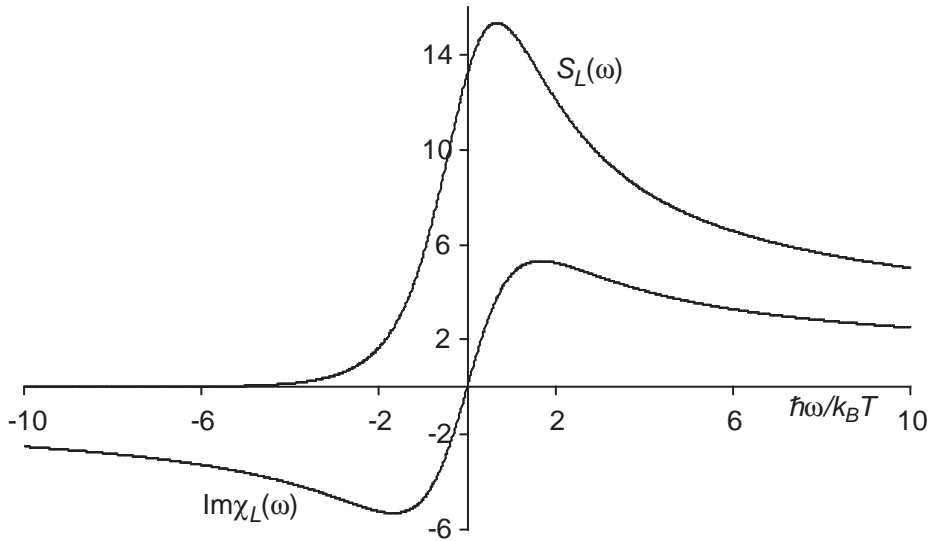}} \caption{Plots of
the dynamic local susceptibility $\mbox{Im} \chi_L (\omega)$, in
(\protect\ref{cl1}), and the local dynamic structure factor
(measured in neutron scattering experiments) $S_L (\omega) = 2
\mbox{Im} \chi_L (\omega)/(1 - e^{-\hbar\omega/k_B T})$ at
$\alpha=1/2$ and some fixed temperature $T$. Note that $\mbox{Im}
\chi_L (\omega)$ is an odd function of $\omega$, while $S_L (
\omega)$ is not an even function. At large $|\omega|$, $\mbox{Im}
\chi_L (\omega) \sim \mbox{sgn}(\omega)/|\omega|^{1-\alpha}$.}
\label{fig4}
\end{figure}

The result (\ref{nn}) allows us to directly compute the observable
local dynamic spin susceptibility, $\chi_L (\omega)$ in the
quantum critical region; we define this quantity by the Fourier
transform in imaginary frequency
\begin{equation}
\chi_L (i \omega_n) = \int_0^{\beta} d \tau e^{i \omega_n \tau}
\langle {\bf n} (\tau) \cdot {\bf n} (\tau') \rangle_{{\cal Z}}.
\label{cl}
\end{equation}
While the result (\ref{nn}) is exact at $T=0$ for the model
obeying (\ref{da}), the form of the exact result at $T>0$ is not
known. However, a solution for the $T>0$ dynamics is possible in
the large $N$ limit\cite{gp,kot}; the right hand side of
(\ref{nn}) is replaced by $[\pi T/\sin(\pi T
|\tau-\tau^{\prime}|)]^{\alpha}$, and its Fourier transform yields
the result
\begin{equation}
\mbox{Im}\left[ \chi_L ( \omega) \right] \sim T^{\alpha-1} \sinh
\left(\frac{\hbar\omega}{2 k_B T} \right) \left| \Gamma \left(
\frac{\alpha}{2} - i \frac{\hbar \omega}{2 \pi k_B T} \right)
\right|^2 \label{cl1}.
\end{equation}
This constitutes the complete description of the relaxational
quantum critical spin dynamics: note again that the function
depends universally on $\hbar \omega/k_B T$. At the important
value $\alpha=1$, (\ref{cl1}) reduces to the simple result
$\mbox{Im}\left[ \chi_L ( \omega) \right] \sim \tanh
({\hbar\omega}/{2 k_B T})$. We show a plot of the spectral
function (\ref{cl1}) at $\alpha=1/2$ in Fig~\ref{fig4}.

The result (\ref{cl1}) is also that expected if the local spin
dynamics obeys a conformally invariant field theory in 1+1
dimensions~\cite{m1}. There is no reason for this to be generally
the case for ${\cal Z}$, and (\ref{cl1}) has been shown to hold
only in a large $N$ limit; we expect corrections to the scaling
function of $\omega/T$ at higher orders, and computing these
remains an important challenge. Comparisons of (\ref{cl1}) with
experimental data on UCu$_{5-x}$Pd$_{x}$ have been
made\cite{m1,m2,m3}; while some higher frequency features are
reasonably captured, the agreement is poor for $\hbar \omega \ll
k_B T$, and it has been proposed that this is due to the effects
of disorder\cite{m3}.

\section{Conclusions}

We have presented here a series of paradigms of dynamic response
functions at finite temperatures near a quantum critical point.
The behavior discussed here is characteristic of critical points
below their upper critical dimension. As the number and precision
of experimental measurements of such critical points increases, we
hope that the accuracy of theoretical predictions will be
eventually be sufficient to permit a quantitative confrontation
between theory and experiment.

We have not discussed here the dynamics of systems above their
upper-critical dimension: in these cases the dynamics usually
remains quasiclassical even near the critical point, and a
van-Hove like theory with a relaxation rate dependent upon
microscopic details provides an adequate description.

\section*{Acknowledgments}
This research was supported by US NSF Grant DMR 0098226.


\begin{thebibliography}{99}

\bibitem{forster} D.~Forster, {\em Hydrodynamic Fluctuations,
Broken Symmetry, and Correlation Functions}, Benjamin Cummings,
Reading, Mass. (1975).

\bibitem{lovesey} S.~W.~Lovesey, {\em Condensed Matter Physics,
Dynamic Correlations}, Benjamin Cummings, Reading, Mass. (1980).

\bibitem{landau} L.~D.~Landau and E.~M.~Lifshitz, {\em Statistical
Physics}, Pergamon Press, Oxford (1980).

\bibitem{halphoh} B.~I.~Halperin and P.~C.~Hohenberg, \RMP~{\bf 49},
435 (1977).

\bibitem{book}S.~Sachdev, {\em
Quantum Phase Transitions}, Cambridge University Press, Cambridge
U.K. (1999).

\bibitem{chn} S.~Chakravarty, B.~I.~Halperin, and D.~R.~Nelson,
\PRB\ {\bf 39}, 2344 (1989).

\bibitem{sy} S.~Sachdev and J.~Ye, \PRL\ {\bf 69}, 2411 (1992).

\bibitem{csy} A.~V.~Chubukov, S.~Sachdev, and J.~Ye, \PRB\ {\bf 49}, 11919
(1994).

\bibitem{ssrelax} S.~Sachdev, \PRB\ {\bf
59}, 14054 (1999).

\bibitem{damle} K.~Damle and S.~Sachdev, \PRB\ {\bf 56}, 8714 (1997).

\bibitem{apy} S. Sachdev and A.P. Young, \PRL\ {\bf 78}, 2220
(1997).

\bibitem{altenberg} S.~Sachdev, cond-mat/0109419.

\bibitem{korepin} V.~E.~Korepin, N.~M.~Bogoliubov, and A.~G.~Izergin,
{\em Quantum Inverse Scattering Method and Correlation Functions,}
Cambridge University Press, Cambridge U.K. (1993).

\bibitem{pfeuty} P.~Pfeuty {\it Ann. of Phys.} {\bf 57}, 79
(1970).

\bibitem{valla} T.~Valla, A.~V.~Fedorov, P.~D.~Johnson,
B.~O.~Wells, S.~L.~Hulbert, Q.~Li, G.~D.~Gu, and N.~Koshizuka,
\Sci\ {\bf 285}, 2110 (1999).

\bibitem{vojta1} M.~Vojta, Y.~Zhang, and S.~Sachdev, \PRB\ {\bf 62}, 6721
(2000).

\bibitem{vojta2} M.~Vojta, Y.~Zhang, and S.~Sachdev, \PRL\ {\bf 85},
4940 (2000).

\bibitem{dagan} Y.~Dagan and G.~Deutscher, \PRL\ in
press, cond-mat/0106128.

\bibitem{metzner} C.~J.~Halboth and W.~Metzner \PRL\ {\bf
85}, 5162 (2000).

\bibitem{maxim1} M.~A.~Baranov and M.~Yu Kagan, \ZPB\ {\bf 86}, 237 (1992).

\bibitem{maxim2} M.~A.~Baranov, A.~V.~Chubukov, and M.~Yu Kagan,
\IJMPB\ {\bf 6}, 2471 (1992).

\bibitem{kvesch} D.~V.~Khveshchenko and J.~Paaske, \PRL\ {\bf 86}, 4672
(2001).

\bibitem{qmc} L.~K\"arkk\"ainen, R.~Lacaze, P.~Lacock, and B.~Petersson,
\NPB\ {\bf 415}, 781 (1994).

\bibitem{piers} P.~Coleman, C.~Pepin, Q.~Si, and R.~Ramazashvili,
\JPC\ {\bf 13}, 723, (2001); P.~Coleman and C.~Pepin,
cond-mat/0110063.

\bibitem{almut} A.~Schr\"oder, G.~Aeppli, R.~Coldea, M.~Adams,
O.~Stockert, H.v.~L\"ohneysen, E.~Bucher, R.~Ramazashvili, and
P.~Coleman, \Nat\ {\bf 407}, 351 (2000).

\bibitem{hertz} J.~A.~Hertz, \PRB\ {\bf 14}, 1165 (1976).

\bibitem{millis} A.~J.~Millis, \PRB\ {\bf 48}, 7183 (1993).

\bibitem{royal} S.~Sachdev, {\em Phil. Trans. Roy. Soc. London}
A {\bf 356}, 173 (1998).

\bibitem{maclaughlin1} D.~E.~MacLaughlin, O.~O.~Bernal, R.~H.~Heffner,
G.~J.~Nieuwenhuys, M.~S.~Rose, J.~E.~Sonier, B.~Andraka, R.~Chau,
and M.~B.~Maple, \PRL\ {\bf 87}, 066402 (2001).

\bibitem{aronson} M.~C.~Aronson, R.~Osborn, R.~A.~Robinson, J.~W.~Lynn,
R.~Chau, C.~L.~Seaman, and M.~B.~Maple, \PRL\ {\bf 75}, 725
(1995).

\bibitem{maclaughlin2} D.~E.~MacLaughlin, R.~H.~Heffner,
G.~J.~Nieuwenhuys, G.~M.~Luke, Y.~Fudamoto, Y.~J.~Uemura, R.~Chau,
M.~B.~Maple, and B.~Andraka, \PRB\ {\bf 58}, R11849 (1998).

\bibitem{sy2} S.~Sachdev and J.~Ye, \PRL\ {\bf 70}, 3339
(1993).

\bibitem{gps} A.~Georges, O.~Parcollet, and S.~Sachdev, \PRL\ {\bf 85}, 840
(2000); \PRB\ {\bf 63}, 134406 (2001).

\bibitem{gp} O.~Parcollet and A.~Georges, \PRB\ {\bf 59}, 5341
(1999).

\bibitem{si} Q.~Si, S.~Rabello, K.~Ingersent, and J.~L.~Smith,
cond-mat/0011477.

\bibitem{anirvan} A.~M.~Sengupta, \PRB\ {\bf 61}, 4041
(2000).

\bibitem{qmsi} J.~L.~Smith and Q.~Si, \EPL\ {\bf 45}, 228 (1999).

\bibitem{vbs} M.~Vojta, C.~Buragohain, and S.~Sachdev,
\PRB\ {\bf 61}, 15152 (2000).

\bibitem{kot} O.~Parcollet, A.~Georges, G.~Kotliar, and A.~M.~Sengupta,
\PRB {\bf 58}, 3794 (1998).

\bibitem{m1} M.~C.~Aronson, M.~B.~Maple, P.~de Sa, A.~M.~Tsvelik,
and R.~Osborn, \EPL\ {\bf 40}, 245 (1997).

\bibitem{m2} M.~C.~Aronson, M.~B.~Maple, R.~Chau, A.~Georges, A.~M.~Tsvelik,
and R.~Osborn, \JPC\ {\bf 8}, 9815 (1996).

\bibitem{m3} R.~Chau, M.~C.~Aronson, E.~J.~Freeman, and
M.~B.~Maple,\JPC\ {\bf 12}, 4495 (2000).

\end{thebibliography}
\end{document}